\documentclass{article}
\usepackage{graphicx}
\usepackage{amssymb, amsmath}
\usepackage{amsfonts}
\usepackage{epstopdf}

\title{Dynamical Wave Function Collapse: Could It Have Cosmological Consequences?}
\author{Philip Pearle\\ 
Hamilton College\\
 Clinton, NY 13323, USA\\
e-mail: ppearle@hamilton.edu}

\begin{document}
\date{}
\maketitle
\abstract{This talk was in response to the conference injunction to present "one bold but half-baked idea that you have been thinking about recently.. .".  After a brief introduction to the ideas and formalism of the CSL theory of dynamical wave function collapse, some possible cosmological effects associated with applying it to the creation of the universe are discussed. }

\section{Why Change Quantum Theory?}
	
	An interpretation of a theory may be defined as a set of rules for going from mathematical statements to statements about reality.  In particular, given an initial state vector and the Hamiltonian governing its evolution, an interpretation of quantum theory should enable one to say which are the states which might be realized in nature, and their probabilities of realization.  
	
	The famous so-called ``measurement problem," which I prefer to call the ``reality problem" is that no currently proposed interpretation of standard quantum theory is well-defined.  The ``Copenhagen Interpretation" rules rely upon the undefined notion of apparatus.  The ``Everett/Relative State/Many Worlds Interpretations" rules rely upon the undefined notion of observer or (in its most recent manifestation) the success of the Decoherent Histories program.  If this latter is successful, it would constitute an interpretation, but it is not at present well-defined in that general rules for picking the projection operators and times of projection which its formalism requires are  still lacking.  
	
	Quantum theory has been around a long time, so one might reasonably suspect that it is incapable of supporting a well-defined interpretation.   If a mathematical theory is not well-defined, that is obviously reason for improving it.  The same should be true of a physical theory.   
	
	This is the motivation for altering quantum theory so that it describes wave function collapse as a dynamical, physical, process.  In the theory discussed here (called Continuous Spontaneous Localization, or CSL), an anti-Hermitian, operator is added to the Hamiltonian in Schr\"odinger's equation. This operator is  the mass density of particles coupled to a c-number randomly fluctuating scalar field $w({\bf x},t)$.  The altered  evolution  evolves a state vector, expressed as a superposition of states of different mass density configurations,  toward one of those states. 
	
	The evolution is very slow  if  e.g., the states differ in the relative displacement of just a few particles, so that the usual quantum theory predictions of microscopic behavior are negligibly affected.  The evolution is very rapid if e.g., the states differ in the relative displacement of a macroscopic object:   thus, the state vector describes the macroscopic world we see around us instead of (what may be thought of as) a superposition of such worlds. 
	
	In addition to the altered Schr\"odinger equation, the theory supplies a second equation, the ``probability rule."  It says that the probability any particular field  $w({\bf x},t)$ (and therefore its associated  state vector) is realized in nature, is proportional to the squared norm of the (non-unitarily evolving) state vector.  So, large norm state vectors are more probable than small norm state vectors.  Applied to  all so-far performed experiments, the probability predictions of CSL differ  undetectably from that given by the Born Rule.  However, there \textit{are} experiments, some of which may soon prove feasible, whose outcomes are predicted to differ from those predicted by standard quantum theory.  
	
	CSL \textit{does} resolve the reality (measurement) problem: it  \textit{has}  a well-defined interpretation.   Put \textit{any} $w({\bf x},t)$ (of white noise type) into the modified of Schr\"odinger equation, and the resulting state is a realizable state of nature.  The probability rule gives its probability of realization.  
	
	Feynman wrote in 1965:
{\vskip 10pt}	
		\footnotesize``We have to find a new view of the world that has to agree with everything that is known, but disagree in its predictions somewhere, otherwise it is not interesting.  And in that disagreement it must agree with nature.  If you can find any other view of the world which agrees over the entire range where things have already been observed, but disagrees somewhere else, you have made a great discovery.  It is very nearly impossible, but not quite, to find any theory which agrees with experiments over the entire range in which all theories have been checked, and yet gives different consequences in some other range, even a theory whose different consequences do not turn out to agree with nature. " 
		{\vskip 10pt}				
	\normalsize \noindent CSL may be considered, at present, to be described by most of the last sentence: whether it turns out to be described by the next-to-last sentence, or just the last phrase of the last sentence, remains to be seen.  
	
\section{How It Works}	
	
	Idealized collapse dynamics works as follows. 	If the initial state vector is 	
\begin{equation}\label{1}
|\psi, 0\rangle=\sum_{n=1}^{N}c_{n}(0)|a_{n}\rangle,  
\end{equation}
(where the $|a_{n}\rangle$ are eigenstates of an operator $A$ with nondegenerate eigenvalues $a_{n}$), it should evolve as $t\rightarrow\infty$ to one of the  $|a_{n}\rangle$ with probability 
$|c_{n}(0)|^{2}$.  

	There is an intuitively appealing analogy to collapse dynamics, the Gambler's Ruin Game.  Consider, for simplicity, two gamblers, one of whom has \$60, the other \$40.  (This is analogous to there being two states $|a_{1}\rangle$, $|a_{2}\rangle$, with initial respective amplitudes $c_{1}(0)=\sqrt{.6}$, $c_{2}(0)=\sqrt{.4}$).  They toss a coin: heads, one gambler receives a dollar from the other, tails, the dollar goes the other way.  (Analogously, the amplitudes $c_{1}(t)$, $c_{2}(t)$ fluctuate with time, but  
$|c_{1}(t)|^{2}+|c_{2}(t)|^{2}=1$).  Eventually one of the gamblers wins all the money so the game stops:  the gambler with \$60 initial stake, will win 60\% of such repeated games. (Analogously,  $|c_{1}(t)|\rightarrow 1$,  $|c_{2}(t)|\rightarrow 0$ for 60\%  of the evolutions).  
	
	Here is the evolution which mimics this game's behavior.   The solution of the modified Schr\"odinger equation mentioned earlier, with the Hamiltonian $H=0$, is	
\begin{eqnarray}\label{2}
|\psi, t\rangle_{w}&\equiv&e^{-(4\lambda)^{-1}\int_{0}^{t}dt'[w(t')-2\lambda A]^{2}}|\psi, 0\rangle\nonumber\\
&=&\sum_{n=1}^{2}c_{n}|a_{n}\rangle e^{-(4\lambda)^{-1}\int_{0}^{t}dt'[w(t')-2\lambda a_{n}]^{2}},   
\end{eqnarray}
 \noindent where the constant $\lambda$ characterizes the collapse rate and the second equation in (2) utilizes Eq. (1) with $N=2$. 
 
  The probability associated to $|\psi, t\rangle_{w}$, as mentioned earlier,  is 
   \begin{equation}\label{3}
P_{w}(t)Dw\equiv  _{w}\negthinspace\negthinspace\negthinspace\langle \psi, t|\psi, t\rangle_{w}Dw=\sum_{n=1}^{2}|c_{n}|^{2} e^{-(2\lambda)^{-1}\int_{0}^{t}dt'[w(t')-2\lambda a_{n}]^{2}}Dw.   
\end{equation}
\noindent In Eq. (3), $Dw\equiv Cdw(0)dw(\Delta t)dw(2\Delta t)...dw(t)$, and $C=(2\pi\lambda/\Delta t)^{-t/\Delta t}$, so that the total probability, integrated over all $w(n\Delta t)$, is 1.  It can readily be shown that only $w({\bf x},t)$'s for which 
\[
 T^{-1}\int_{0}^{T}dt'w(t')\rightarrow 2\lambda a_{1} \hbox {\quad or} \rightarrow 2\lambda a_{2} \hbox{\quad as\quad} T\rightarrow \infty
\]
have non-vanishing asymptotic probability (3).  If e.g.,  $T^{-1}\int_{0}^{T}dt'w(t')\rightarrow 2\lambda a_{1}$, Eqs. (2), (3) asymptotically become
\begin{equation}\label{4}
|\psi, t\rangle_{w}\approx c_{1}|a_{1}\rangle e^{-(4\lambda)^{-1}\int_{0}^{t}dt'[w(t')-2\lambda a_{1}]^{2}},   
\end{equation}	  
 \begin{equation}\label{5}
P_{w}(t)Dw\approx |c_{1}|^{2} e^{-(2\lambda)^{-1}\int_{0}^{t}dt'[w(t')-2\lambda a_{1}]^{2}}Dw.    
\end{equation}	  	    	 
\noindent 	 (4) is a (un-normalized) collapsed state, and the integral of  (5)'s probability over all $w(t)$'s is $ |c_{1}|^{2}$.

The density matrix constructed from  (2), (3) is 
\begin{equation}\label{6}
\rho=\int P_{w}(t)Dw\frac{|\psi, t\rangle_{w}\thinspace_{w}\langle \psi,t|}{_{w}\langle \psi,t|\psi,t\rangle_{w}}=\sum_{n, m=1}^{2}c_{n}c_{m}^{*}|a_{n}\rangle\langle a_{m} |e^{-(\lambda t/2)(a_{n}-a_{m})^{2}},
\end{equation}
\noindent from which one can see that the decay rate of the off-diagonal elements  increases as the eigenvalue difference increases.  

	For many mutually commuting operators $A_{k}$, and with a possibly time-dependent Hamiltonian $H(t)$, the evolution (2) becomes 
\begin{equation}\label{7}
	|\psi, t\rangle_{w}\equiv {\cal T}e^{-\int_{0}^{t}dt' \{iH(t')+(4\lambda)^{-1}\sum_{k}[w_{k}(t')-2\lambda A_{k}]^{2}\}}|\psi, 0\rangle  
\end{equation}
\noindent (${\cal T}$ is the time-ordering operator).
	
 	For CSL,  the index $k$ corresponds to spatial position ${\bf x}$, so that $w_{k}(t)\rightarrow w({\bf x}, t)$ can be regarded as a physical field.   $A_{k}\rightarrow A({\bf x})$ is chosen to be the mass  density operator $M({\bf x})$ ``smeared" over a region of length $a$ (a second parameter in the theory) around ${\bf x}$:    
\begin{equation}\label{8}
	|\psi, t\rangle_{w}\equiv {\cal T}e^{-\int_{0}^{t}dt' \{iH(t')+(4\lambda)^{-1}\int d {\bf x}[w({\bf x}, t')-2\lambda A({\bf x})]^{2}\}}|\psi, 0\rangle,  
\end{equation}	
\begin{equation}\label{9}
\frac{d\rho(t)}{dt}=-i[H,\rho(t)]-\frac{\lambda}{2}\int d{\bf x}[A({\bf x}),[A({\bf x}), \rho(t)]]
\end{equation}
\begin{equation}\label{10}
A({\bf x})\equiv\frac{1}{m_{0}(\pi a^{2})^{3/4}}\int  d {\bf z} e^{-\frac{1}{2a^{2}}({\bf x}-{\bf z})^{2}}M({\bf z}),     
\end{equation}
\noindent ($m_{0}$ is taken to be the proton's mass).  Eq.(9) follows from (8) and the first equations in  (3), (6).  The parameter values $\lambda=10^{-16}$sec$^{-1}$ and $a=10^{-5}$cm, which were chosen by Ghirardi, Rimini and Weber for their  collapse model  (which contributed essential ideas to CSL) shall be adopted.  If CSL indeed describes nature, these numbers should be open to experimental determination.  

	The dynamical equation (8) and the probability rule (the first equation in (2) constitute the CSL model, which can be applied to any non-relativistic physical system.  CSL works by recognizing a superposition of states which differ in their distribution of mass density, and conducting a gambler's ruin-type competition among them. 

\section{A  Role For Collapse in Cosmogenesis?}
	The instigators of this workshop requested that we speak about ``One bold but half-baked (or half- to three-quarters-baked) idea that you have been thinking about recently... ."  Here's one.  If the beginning of our universe was a quantum event, and if state vector collapse is a real physical process,  then perhaps it played a role in the selection, and even in the generation, of our universe.   
	
	Here is a simple illustrative  model.  One might suppose that there is a Hamiltonian which describes the creation of the universe out of the vacuum.  The  Schr\"odinger evolution might produce a superposition of different geometries, of different pre-inflationary configurations, etc.  For our much less sophisticated model, we shall take a Hamiltonian, acting in a pre-ordained volume $V$, which produces a superposition of different numbers of particles of mass $m$ out of the initial vacuum state $|0\rangle$,  and which acts for a time interval $T$:
\begin{equation}\label{10}
H=\int_{V} d{\bf x}\{m\xi^{\dagger}({\bf x})\xi({\bf x})+g[\xi({\bf x})+\xi^{\dagger}({\bf x})]\}.
\end{equation}
\noindent where  $\xi({\bf x})$ is the annihilation operator for a particle at ${\bf x}$ and $g$ is a coupling constant.  $H$ can be thought of as describing a displaced harmonic oscillator at every point of space.   The solution of the usual Schr\"odinger equation is 
\begin{equation}\label{11}
|\psi, t\rangle=e^{iTg^{2}V/m}e^{-(g/m)\int_{V} d{\bf x}[\xi^{\dagger}({\bf x})+(g/m)]
[1-e^{-imT}]}|0\rangle.
\end{equation}
The mean number of particles oscillates:  
\begin{equation}\label{12}
\overline n(T)\equiv\langle \psi,t|\int_{V} d{\bf x}\xi^{\dagger}({\bf x})\xi({\bf x})|\psi, t\rangle=2V\bigg(\frac{g}{m}\bigg)^{2}(1-\cos mT).
\end{equation}

	Now, suppose that CSL collapse dynamics holds even at the beginning of the universe.  Using Eq. (9) , one may  find coupled equations for $\overline n(T)$ (here defined as $\overline n(T)\equiv$Tr$\int_{V} d{\bf x}\xi^{\dagger}({\bf x})\xi({\bf x})\rho(T)$) and Tr$\int_{V} d{\bf x}[\xi^{\dagger}({\bf x})\pm\xi({\bf x})]\rho(T)$ (Tr is the trace operator), with the result
\begin{equation}\label{13}
\overline n(T)=\frac{g^{2}V}{m^{2}+(\lambda/2)^{2}}\{\lambda T-2[\cos\theta-e^{-(\lambda/2)T}\cos(\theta+mT)]\}
\end{equation}
($\theta\equiv2\tan^{-1}(2m/\lambda))$.  Two things are happening here.  

One is that, as the Hamiltonian generates a superposition of different ``universes" with different particle density distributions, the collapse term acts to select one of them, ``our universe."  

The other is that the collapse effectively continually excites the harmonic oscillators at each point of space.  The result, on average, is that the number of particles in the universe grows linearly with time,  $\sim g^{2}\lambda$.  That is, the creation of an interesting universe is a cooperative venture, requiring \textit{both} Hamiltonian and collapse dynamics.  

(As an aside, note that $\overline n(T)\rightarrow 0$ as $\lambda\rightarrow \infty$.  The universe remains in the vacuum state due to ``watched pot" or ``Zeno's paradox" behavior: the collapse occurs so fast that there is no chance for the vacuum state to evolve).  
\section{A Gravitational Role For Collapse?}
	Collapse dynamics narrows particle wave functions and so, by the uncertainty principle, particles gain energy.  For example,  there is a small probability that an electron will be knocked out of an atom, or a nucleon out of a nucleus.  
	
	One can show from (9), regardless of the interaction potential, that the ensemble average energy $E\equiv$Tr$H\rho(t)$ gained by each particle of mass $m$ over the age $t$ of the universe is  
\begin{equation}\label{14}	
E=t\frac{3\hbar^{2}\lambda m}{(2m_{0}a)^{2}}\approx10^{-16}mc^{2}.
\end{equation}	
This is a very small amount, certainly not of cosmological significance.  Nonetheless, conservation of energy is an important physical principle, and it would be good not to violate it.  It turns out that one can associate an energy with $w({\bf x}, t)$, and with its interaction with particles, such that the total energy is conserved. This involves quantizing the $w$-field, but we shall not need the details here.  

	As is usual in quantum theory, conservation of energy is guaranteed to hold for the  whole ensemble, not for the individual states in the ensemble.  However, as in quantum theory, there are certain circumstances where energy conservation does hold for individual states.  For example, expand the initial state in eigenstates of the particle density basis.   If, in this basis, off-diagonal matrix elements of powers of  $H$ vanish (as would be the case if these basis states are macroscopically distinct), then the energy associated with each such evolving basis state is separately conserved.  That is, the energy spectrum of the initial basis state is the same as the energy spectrum of all the collapsed states to which this initial basis state evolves.   
	
	For our half-baked application, lets concentrate just upon the expectation values of the particle energy, the $w$-field energy,  and their energy densities. The total energy is the sum of the particle energy, the $w$-field energy and the particle-$w$ interaction energy.  It can be shown that the expectation value of the  interaction energy vanishes.  Thus, the sum of the  expectation value of the particle energy and the $w$-field energy is constant.  Assume that the $w$-field energy's initial value is zero.  As the dynamics progresses,  its expectation value becomes negative, since the expectation value of the particle energy increases.  The expectation value of the $w$-field energy \textit{density}, starting at 0, can go positive or negative, but its integral over all space is negative. The $w$-field energy density in this formulation doesn't move:  once it is created in a volume, it just stays stuck in space.  Further particle collapse in that volume subtracts or adds to the energy density there, such that the expectation value of the integrated energy density continues to diminish.  
			
	Now for the half-baked idea.  Consider semi-classical gravity, the Einstein equation's left side equated to the quantum expectation value of the stress tensor of the particles in the universe.  This cannot be a correct equation if there is collapse dynamics. The covariant divergence of the left side of the equation vanishes, but the covariant divergence of the right side does not, since the particles alone do not conserve energy-momentum.  This argues that, to the  stress tensor term belonging to the particles, there ought to be added a stress tensor term belonging to the $w$-field, such that the divergence of the sum vanishes. 
	
	Given such a $w$-field stress tensor term (full half-baked disclosure---I do not have such a term), it then follows that the $w$-field energy exerts a gravitational force.  In particular, if the energy density is negative, it repels particles.  
	
	The putative negative energy the $w$-field acquires, equal to the kinetic energy (15) gained by particles undergoing collapse is, likewise,  much too small to be of cosmological significance.  However, it is otherwise for the universe-creation scheme of the previous section.   
	
	As the universe grows from the vacuum and acquires particle mass-energy, the $w$-field acquires a comparable negative energy.  Thus, the picture emerges of a $w$-field density, positive in some places, negative in others, such that the total $w$-field energy is negative and equal to the total particle mass-energy in the universe.  Once creation of the universe ceases, the $w$-field energy density, positive and negative, is stuck in space, if we accept the behavior mentioned previously.  One might  expect that the negative energy density would repel matter, creating voids, while the positive energy density would attract matter,  perhaps acting as seeds for galaxy formation.  
	
	To conclude, my question at this workshop was, might this set of ideas, or some variation on them, be of any use to astrophysical modeling?   What might be the effect of the $w$-field energy on the expansion of the universe?  Might such ideas be testable or ruled out by observation? 
	   
\end{document}